\documentclass[article]{biometrika}

\usepackage{amsmath}

\usepackage{times}
\usepackage{bm}
\usepackage{natbib}

\usepackage[plain,noend]{algorithm2e}

\makeatletter
\renewcommand{\algocf@captiontext}[2]{#1\algocf@typo. \AlCapFnt{}#2} 
\def\@algocf@capt@plain{top}
\renewcommand{\algocf@makecaption}[2]{%
	\addtolength{\hsize}{\algomargin}%
	\sbox\@tempboxa{\algocf@captiontext{#1}{#2}}%
	\ifdim\wd\@tempboxa >\hsize
	\hskip .5\algomargin%
	\parbox[t]{\hsize}{\algocf@captiontext{#1}{#2}}
	\else%
	\global\@minipagefalse%
	\hbox to\hsize{\box\@tempboxa}
	\fi%
	\addtolength{\hsize}{-\algomargin}%
}
\makeatother

\newcommand{\h}	{\mathcal{H}}

\newcommand\numberthis{\addtocounter{equation}{1}\tag{\theequation}}

\usepackage{tikz}
\usetikzlibrary{arrows,shapes.arrows,shapes.geometric,
	shapes.multipart,backgrounds,decorations.pathmorphing,positioning,fit,automata}
\tikzset{
	>=stealth',
	true/.style={
		rectangle,
		draw=black, very thick,
		text width=6.5em,
		minimum height=2em,
		text centered,
		fill=gray, opacity = 0.5},
	punkt/.style={
		rectangle,
		rounded corners,
		draw=black, very thick,
		text width=6.5em,
		minimum height=2em,
		text centered},
	est/.style={
		circle,
		draw=black, very thick,
		text centered},
	shade/.style={
			circle,
			draw=black, very thick, fill=gray!50,
			text centered},
	weight/.style={
		circle,
		draw=black, very thick,
		text width=6.5em,
		minimum height=2em,
		text centered},
	pil/.style={
		->,
		thick,
		shorten <=2pt,
		shorten >=2pt,},
	double/.style={
		<->,
		thick,
		shorten <=2pt,
		shorten >=2pt,},
	dash/.style={
		dashed,
		thick,
		shorten <=2pt,
		shorten >=2pt,},
	dashdouble/.style={
		<->,
		dashed,
		thick,
		shorten <=2pt,
		shorten >=2pt,}
}
\usepackage{etex}
\usepackage{booktabs}
\usepackage{amsfonts}

\begin{document}
	
	\jname{Biometrika}
	\jyear{}
	\jvol{}
	\jnum{}
	
	
	\markboth{L. Wang, J. M. Robins \and T. S. Richardson}{Miscellanea}
	
	\title{On {falsification of} the binary instrumental variable model}
	
\author{LINBO WANG}
\affil{Department of Biostatistics, Harvard School of Public Health,
	677 Huntington Avenue, Boston, Massachusetts 02115, U.S.A.
 \email{linbowang@g.harvard.edu }}

\author{ JAMES M. ROBINS}
\affil{Department of Epidemiology, Harvard School of Public Health,
	677 Huntington Avenue, Boston, Massachusetts 02115, U.S.A.
	\email{robins@hsph.harvard.edu}}

\author{\and THOMAS S. RICHARDSON}
\affil{Department of Statistics, University of Washington, Box 354322, Washington 98195,
	U.S.A.
	\email{thomasr@u.washington.edu}}
	
	\maketitle
	
	
\begin{abstract}
	Instrumental variables are widely used for estimating  causal effects in the presence of unmeasured confounding. The discrete instrumental variable model has testable implications on the law of the observed data. However, current assessments of instrumental validity are typically based solely on subject-matter arguments rather than these testable implications, partly due to a lack of formal statistical tests with known properties. In this paper, we develop simple procedures for testing the binary instrumental variable model. Our methods are based on existing approaches for comparing two treatments, such as the t-test and the Gail--Simon test. We illustrate the importance of testing the instrumental variable model by evaluating the exogeneity of college proximity using the National Longitudinal Survey of Young Men. 
\end{abstract}


\section{Introduction}
%

\begin{figure}[!htbp]
	\centering
	\begin{tikzpicture}[->,>=stealth',node distance=1cm, auto,]
	\node[est] (Z) {$Z$};
	\node[est, right = of Z] (D) {$D$};
	\node[est, right = of D] (Y) {$Y$};
	\node[shade, below = of D] (U) {$U$};
	\node[est, above = of D] (V) {$V$};
	\path[pil] (Z) edgenode {} (D);
	\path[pil] (D) edgenode {} (Y);
	\path[pil] (U) edgenode {} (D);
	\path[pil] (U) edgenode {} (Y);
	\path[pil] (V) edgenode {} (Z);
	\path[pil] (V) edgenode {} (D);
	\path[pil] (V) edgenode {} (Y);
	\end{tikzpicture}
	\caption{Direct acyclic graph representing an instrumental variable model. The variables $V$, $Z$, $D$ and $Y$ are observed; $U$ is unobserved. }
	\label{DAG:iv_model}
\end{figure}
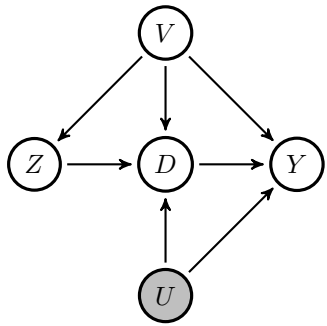

The instrumental variable  method has been widely used for estimating causal effects in  the presence of unmeasured confounders. A variable $Z$ is called an instrumental variable if  (a) it is independent of unmeasured confounders $U$; (b) it does not have a direct effect on the outcome $Y$; (c) it has a non-zero average causal effect on the treatment $D$ \citep{angrist1996identification}. In many  applications  assumption (a) is  reasonable only after controlling for observed covariates $V$ \citep{baiocchi2014instrumental}.  The resulting model is called the {conditional} instrumental variable model. Figure \ref{DAG:iv_model} gives a directed acyclic graphical model representation \citep{pearl2009causality} of the conditional instrumental variable model, in which the faithfulness \citep{spirtes2000causation} of the edge $Z\rightarrow D$ is assumed.

Unlike the assumption of no unmeasured confounders between $D$ and $Y$, the instrumental variable model with discrete observables $(Z,D,Y)$ imposes non-trivial constraints on the observed data distribution. In particular, \cite{balke1997bounds}  and \cite{bonet2001instrumentality} give the following necessary and sufficient condition for an observed data distribution $p(d,y\mid z)$ to be compatible with an unconditional binary instrumental variable model where $Z$, $D$ and $Y$ take values $0$ and $1$:
\begin{equation}
\label{eqn:test_implication}
{\rm pr} (D=d,Y=y\mid Z=1) + {\rm pr}(D=d,Y=1-y\mid Z=0) \leq 1, \quad d=0,1, y=0,1.
\end{equation}
Here {the unconditional instrumental variable model refers to the model with an empty control variable set $V$}.
In particular, if the potential instrument $Z$ is randomized so that assumption (a) holds, then violation of each inequality in \eqref{eqn:test_implication} corresponds to a non-zero average controlled direct effect  of $Z$ on $Y$, which  violates assumption (b) \citep{cai2008bounds,richardson2011transparent}. Although assumption (c)  imposes the following constraint on the observables:
\begin{equation}
\label{eqn:test_3}{\rm pr}(D=1\mid Z=1) \neq {\rm pr}(D=1\mid Z=0),
\end{equation}
 it is in general not possible to reject \eqref{eqn:test_3} with a statistical test. Hence hereafter we do not discuss constraint \eqref{eqn:test_3}.
Similarly, the testable implications of a conditional binary instrumental variable model are given by
\begin{flalign*}	
	 {\rm pr}(D=d,Y=y\mid Z=1,V=v) +& {\rm pr}(D=d,Y=1-y\mid Z=0, V=v) \leq 1, \\
	& \quad \quad \quad \quad \quad \quad \quad \ \  \ d = 0,1, y = 0,1, v \in \mathcal{V}, \numberthis\label{eqn:test_implication2}
	\end{flalign*}
where $\mathcal{V}$ contains all possible values for $V$.
In practice, inequalities \eqref{eqn:test_implication}  can be used to partially test the binary unconditional or conditional instrumental variable model.  Likewise \eqref{eqn:test_implication2} can be used to test the conditional instrumental variable model. In contrast it is impossible to empirically falsify the assumption of no unmeasured confounders between $D$ and $Y$ as in an observational study without an instrument.

Although there have been many discussions on estimation of causal effects under the binary instrumental variable model  \citep{vansteelandt2011instrumental,clarke2012instrumental}, less attention has been drawn to testing validity of the binary instrumental variable model.  
 Prior to our work, \cite{ramsahai2011likelihood} consider testing an unconditional binary instrumental variable model using a likelihood ratio test.  Their approach involves solving a constrained  optimization problem and cannot be used to test the {conditional} binary instrumental variable model as described in Fig. \ref{DAG:iv_model}. Furthermore, their approach tests the four inequalities in \eqref{eqn:test_implication} jointly. {Hence without modification, it can only be used to falsify the binary instrumental variable model, but cannot identify which  specific average controlled direct effect of $Z$ on $Y$ must be positive or negative. }
In a related work, \cite{kang2013causal}  provide a falsification test for the instrumental variable assumptions given knowledge of a subpopulation where the edge $Z \rightarrow D$ is  absent.
{This paper develops a novel {perspective} on falsification of the binary instrumental variable model.} Specifically, we show that 
testing \eqref{eqn:test_implication} or \eqref{eqn:test_implication2} is equivalent to testing for a non-positive effect of the instrument $Z$ on a constructed variable.  


\section{Tests for the unconditional binary instrumental variable model}
\label{sec:unconditional}

	To fix ideas, we first consider testing the instrumental variable inequality
	\begin{equation}
		\label{eqn:iv_ineq}
			{\rm pr}(D=0, Y=1 \mid Z=1) + {\rm pr}(D=0, Y=0 \mid Z=0) \leq 1.
	\end{equation}
	Equation \eqref{eqn:iv_ineq} can be rewritten as
	\begin{equation*}
		{\rm pr}(D=0, Y=1 \mid Z=1) - 1 + {\rm pr}(D=0, Y=0 \mid Z=0) \leq 0.
		\end{equation*}
	Define a new variable
	\[
	Q^{01} \equiv \left\{\begin{array}{cc}
	{I}(D=0, Y=1), & \text{\quad  } Z=1,\\[6pt]
	1-{I}(D=0,  Y=0), &  \text{\quad  } Z=0,
	\end{array}
	\right.
	\]
	where ${I}(\cdot)$ is the indicator function.
	It then follows that
\begin{flalign*}
			{\rm pr}(D=0,Y=1 \mid Z=1) - \{1- {\rm pr}(D=0, & Y=0 \mid Z=0)\} \\ &= {\rm pr}(Q^{01}=1 \mid Z=1) - {\rm pr}(Q^{01}=1 \mid Z=0) \equiv \Delta^{01}.
\end{flalign*}
Testing \eqref{eqn:iv_ineq} is hence equivalent to the testing problem 
\begin{equation}
\label{eqn:test_2by2}
	\mathcal{H}_0^{01}:  \Delta^{01} \leq 0 \text{ \quad vs \quad} 	\mathcal{H}_a^{01}:  \Delta^{01}  > 0,
\end{equation}
which is  simply one-sided testing for a $2\times 2$ table. 

	In general, we have four inequalities of the form \eqref{eqn:iv_ineq} with a binary instrumental variable model, so multiplicity adjustment is needed.    Suppose for now that we have one-sided tests $\phi^{00},\phi^{01},\phi^{10},\phi^{11}$ such that the size of $\phi^{dy}$ goes to 0 asymptotically in the interior of the null space defined by $\h^{dy}_0$. Furthermore, assume that the rejection region of $\phi^{dy}$ has no intersection with the null space defined by $\h^{dy}_0$ \citep{perlman1999emperor}. To get a level-$\alpha$  test for \eqref{eqn:test_implication},  a naive Bonferroni correction would require that each of $\phi^{dy}$ has size less or equal to  $\alpha/4$ for testing $\h^{dy}_0$. 	However, the left-hand sides of the four inequalities in \eqref{eqn:test_implication} sum to 2, and hence at most two of them can hold with equality simultaneously. Based on this, we now show that  it suffices to control the level of each test $\phi^{dy}$ at $\alpha/2.$  
	
\begin{figure} [!htbp]
		\centering
		\vspace{-1cm}
		\includegraphics[width=.6\textwidth]{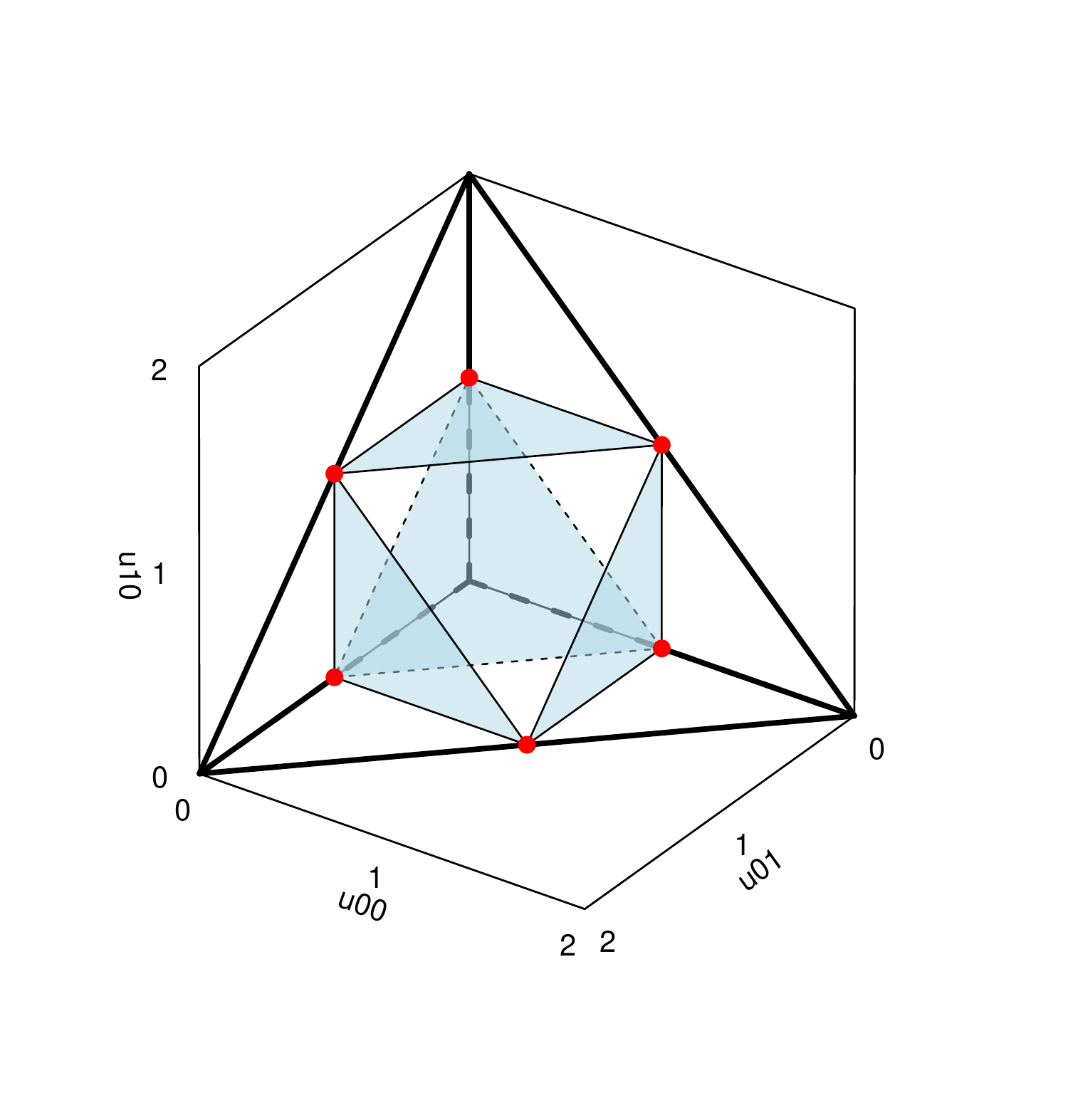}
		\vspace{-1cm}
		\caption{Representation of the simplex $\mathcal{Z}$ and the null space $\mathcal{Z}_0$. The  edges of simplex $\mathcal{Z}$ are annotated using thick black lines, and the null space $\mathcal{Z}_0$ is the octahedron whose vertices are annotated in red, {and four of the eight surfaces are annotated in blue.} }
		\label{fig:octahedron}
	\end{figure}

Specifically, let $u^{dy}={\rm pr}(D=d,Y=y\mid Z=1) + {\rm pr}(D=d, Y=1-y\mid Z=0),$ and $\zeta = (u^{00},u^{01},u^{10})$. The null space defined by \eqref{eqn:test_implication} can be represented by an octahedron $\mathcal{Z}_0$ in the simplex $\mathcal{Z}$, where $\mathcal{Z}$  is defined as 
	$$
			\mathcal{Z} = \left\{\zeta:u^{00}+u^{01}+u^{10} \leq 2, u^{00},u^{01},u^{10} \geq 0 \right\}.
	$$
	Figure \ref{fig:octahedron} gives a graphical depiction of $\mathcal{Z}$ and $\mathcal{Z}_0$.  
Each of	the four blue shaded facets  corresponds to one inequality in \eqref{eqn:test_implication} holding with equality. Six points, annotated in red, have two inequalities in \eqref{eqn:test_implication} holding with equality. The interior of the null space $\mathcal{Z}_0$ corresponds to cases where none of the four inequalities in \eqref{eqn:test_implication} holds with equality.

	We are now ready to present our multiplicity adjustment procedure. The proof of Theorem \ref{thm:main} is given in the Appendix.
	
	\begin{theorem}
		\label{thm:main}
		Propose a testing procedure as follows: reject \eqref{eqn:test_implication} if  for $d=0,1, y=0,1,$ at least one of $\h_0^{dy}$  is rejected by $\phi^{dy}$  at level $\alpha/2$. Under the  null hypotheses \eqref{eqn:test_implication}, 
		\begin{enumerate}[(1)]
			\item if two inequalities in \eqref{eqn:test_implication} hold with equality at  the true value $\dot\zeta$, then the proposed test has size  $\alpha$;
			\item if only one of the inequalities in \eqref{eqn:test_implication} holds with equality at the true value $\dot{\zeta}$, then asymptotically the proposed test has  size  $\alpha/2$;
			\item if none of the inequalities in \eqref{eqn:test_implication} holds with equality at the true value $\dot{\zeta}$, then asymptotically the proposed test has size 0.
		\end{enumerate}
		In particular, the proposed test always has asymptotic size no larger than $\alpha$.
	\end{theorem}

We now turn to choice of $\phi^{dy}$. 
Over the past century, there has been much discussion on testing association in  $2\times 2$ tables{, including size and power comparisons for different test statistics and methods of computing the p-value}; see \cite{lydersen2009recommended} for a review. When the sample size is large,  asymptotic tests 
	such as those based on the t-statistic 
	 are popular among researchers. However, under independent and identically distributed sampling they might not preserve the  test size with small samples, in which case  unconditional exact tests such as the Fisher--Boschloo test are recommended.

	\begin{remark}
		Computation time for unconditional tests can be excessive when the sample size is moderate or large, in which case it may be desirable to use  \cite{berger1994p}'s procedure to reduce computation time. The proposed test still has asymptotic size no larger than $\alpha$ as long as $\gamma < \alpha/2$, where $100(1-\gamma)\%$ is the confidence level for the nuisance parameter.
	\end{remark}

\begin{remark}
The Wald test for the $2\times2$ table corresponding to \eqref{eqn:test_2by2} coincides with the Wald test  for  \eqref{eqn:iv_ineq}, for which  	${\rm pr}(D=0, Y=1 \mid Z=1)$ and ${\rm pr}(D=0, Y=0 \mid Z=0)$ are estimated via maximum likelihood. However, our introduction of $Q^{dy}$ builds the connection between testing unconditional instrumental inequalities and testing $2\times 2$ tables, and hence motivates many more approaches for testing unconditional instrumental inequalities. 
\end{remark}

We now discuss the interpretation of results from our testing procedure.
  As noted by \cite{cai2008bounds} and \cite{richardson2011transparent}, under the randomization assumption, the average controlled direct effect of $Z$ on $Y$, $\textsc{ACDE}(d) = E\{Y(z=1,d=d)\} - E\{Y(z=0,d=d)\}$ satisfies
	\begin{flalign*}
	& {\rm pr}(D=d,Y=1\mid Z=1)	+ 		{\rm pr}(D=d,Y=0\mid Z=0)  - 1  \leq \textsc{ACDE} (d)   \\
	& \quad \leq 1 - 	{\rm pr}(D=d,Y=0\mid Z=1) - {\rm pr}(D=d,Y=1\mid Z=0). \numberthis\label{eqn:acde-iv}
	\end{flalign*}
	It follows that violation of each inequality in \eqref{eqn:test_implication} corresponds to a
	non-zero average controlled direct effect of Z on Y.
	Our testing procedure is hence interpretable in the sense that if we reject the binary instrumental variable model, we would also know which average controlled direct effect is positive or negative. For example, suppose we reject the null that 	${\rm pr}(D=0, Y=1 \mid Z=1) + 		{\rm pr}(D=0, Y=0\mid Z=0) \leq 1$, then from \eqref{eqn:acde-iv}, we would also conclude that $\textsc{ACDE}(0)$ is positive.

\section{Tests for the conditional binary instrumental variable model}
\label{sec:conditional}

Suppose now we wish to test  the instrumental variable inequality that
\begin{equation}
\label{eqn:iv_ineq_cond}
 {\rm pr}(D=0, Y=1 \mid Z=1, V=v) + {\rm pr}(D=0, Y=0 \mid Z=0, V=v) \leq 1, \quad \text{for all } v \in \mathcal{V}.
\end{equation}	
Using the same arguments as in Section \ref{sec:unconditional}, we can rewrite the testing problem of \eqref{eqn:iv_ineq_cond} as
		\begin{equation}
		\label{eqn:h0all}
		\h_{0,c}^{01}:	\text{for all } v \in \mathcal{V},	\Delta^{01}(v) \leq 0  \quad \text{vs} \quad \h_{a,c}^{01}: 	\text{there exists } v \in \mathcal{V},	\Delta^{01}(v) > 0,
		\end{equation}		
		where $\Delta^{01}(v) = {\rm pr}(Q^{01}=1\mid Z=1,V=v) - {\rm pr}(Q^{01}=1\mid Z=0,V=v)$ and letter $c$ in the subscript is short for conditional.
	
	
{Testing problem \eqref{eqn:h0all} concerns the null hypothesis that a particular treatment is at least as good as the other treatment in all subsets of units, which has been studied extensively.}
For example, with  $V$ discrete, the Gail--Simon test for qualitative interaction can be used to test hypotheses of the form \eqref{eqn:h0all} with a slight modification \citep[][p.364]{gail1985testing}. \cite{chang2015nonparametric} consider the problem with a general $V$ based on $\ell_1$-type functionals of uniformly consistent nonparametric kernel estimators of $\Delta^{01}(v)$.
These tests make no assumptions on the functional form of ${\rm pr}(Q^{01}=1\mid Z=z, V=v)$. This is particularly appealing as  $Q^{01}$ is not directly interpretable. 

As we have four hypotheses of the form \eqref{eqn:iv_ineq_cond}, a multiplicity adjustment is warranted. However, unlike the case with unconditional instrumental variable model, the four inequalities in \eqref{eqn:test_implication2} can be violated simultaneously as each of them concerns multiple covariate values. {In other words, no  result analogous to Theorem \ref{thm:main} holds unless $V$ takes only one value.} Instead,  a naive Bonferroni correction may be used to account for multiple comparisons so that to get an overall level-$\alpha$ test, hypotheses of the form \eqref{eqn:iv_ineq_cond} are tested at level $\alpha/4$.

\begin{remark}
	\label{remark:2}
	{When $V$ is discrete, one can alternatively  apply Theorem \ref{thm:main} to test  the following hypotheses for each $v$:
\begin{equation}
\label{eqn:jamie}
	{\rm pr} (D=d,Y=y\mid Z=1, V=v) + {\rm pr}(D=d,Y=1-y\mid Z=0, V=v) \leq 1,  \quad d=0,1, y=0,1, 
	\end{equation}
		and then use a Bonferroni correction to account for multiple testing due to levels of $V$.  In this way, each hypothesis in \eqref{eqn:jamie} is tested at level $\alpha/(2K)$, where $K$ is the number of possible levels for $V$. Since tests of the form \eqref{eqn:iv_ineq_cond} and \eqref{eqn:jamie} are different,  neither approach generally dominates the other. }
\end{remark}

\begin{remark}
	\label{remark:3}
	The Gail--Simon test examines hypotheses \eqref{eqn:test_implication2} for all possible values of $V$ simultaneously. Alternatively, it may be tempting to test \eqref{eqn:test_implication2} for different levels of $V$  and 
	claim $Z$ is a valid instrument within the subset of population 
 for which \eqref{eqn:test_implication2} are not rejected. Failure to  violate an instrumental variable inequality, however, does not prove that $Z$ is an instrument. This will  ultimately
	rest on whether, based on subject-matter knowledge, we believe that we have measured enough covariates $V$ in order to control confounding, as well as subject-matter arguments for the absence of direct effects of $Z$ on $Y$.
%
%
%
	Consequently
	 one should avoid using the test \eqref{eqn:test_implication2}  as a way to restrict the range of $V$, 
	unless a substantive argument could be made as to why $Z$ is  an instrument
	for one range of $V$, but not for another. 
	 
	
\end{remark}

\section{The causal effect of education on earnings}
\label{sec:data}

We illustrate the use of the proposed tests by examining the instrumental variable model assumed by  \cite{okui2012doubly}.  The goal of their analysis is to estimate the causal effect of education on earnings. To account for unobserved preferences for education levels, \cite{okui2012doubly} follow \cite{card1995using} to use  presence of a nearby four-year college  as an instrument. 
 The validity of this approach relies on the assumption that college proximity only affects earnings through education, and conditional on adjusted potential confounders, college proximity is independent of underlying factors that also affect earnings. These assumptions, however, are hardly watertight.  In fact, as pointed out by \cite{card1995using}, living near a college may influence earnings through higher elementary and secondary school quality, and it may also be associated with higher motivation to achieve labor market success.  

To investigate the possible exogeneity of college proximity, we use the data set provided by \cite{okui2012doubly}, which contains 3010 observations from the National Longitudinal Survey  of Young Men. Following \cite{tan2006regression}, we consider education after high school as the treatment $D$. The outcome wage is dichotomized at its median. For illustrative purposes, we consider three instrumental variable models with nested sets of covariates: (I)  experience only; (II) experience and  race; (III) experience, race and region of residence. The third set was also considered previously by \cite{okui2012doubly}. We use the Gail--Simon test with Bonferroni correction to examine the testable implications of these instrumental variable models.   

\begin{table}[!htbp]
	\begin{center}
		\caption{P-values  and number of subgroups from partial tests for the binary instrumental variable models using college proximity as an instrument for education after high school}
		\label{tab:401k}
\begin{tabular}{lcccccccccc}
\\
	Covariate set $V$ & $\h_{0,c}^{00}$  & $\h_{0,c}^{01}$    & $\h_{0,c}^{10}$   &  $\h_{0,c}^{11}$ & No. of subgroups \\[5pt]
	(I)  &  1.000  & 0.010 &  1.000  &  0.034  & 24  \\
	(II) &  1.000  & 0.132 & 1.000 & 0.143 & 47  \\
	(III)  &  1.000 & 1.000 & 1.000 & 1.000  & 819  \\
\end{tabular}
\label{result}
\end{center}
\end{table}

Table \ref{tab:401k} summarizes the test results. 
The model conditional only on experience is rejected by the proposed test. The p-value from the test on $\h_{0,c}^{01}$ is  significant at the 0.05 level, and the p-value from the test on $\h_{0,c}^{11}$ is also borderline significant. These show that either college proximity has  positive {direct} effects on earnings in some subgroups, or after adjusting for experience college proximity is still correlated with unmeasured confounders such as underlying motivation for labor market success. 
The proposed test fails to reject \cite{okui2012doubly}'s instrumental variable model.  {However, as we discussed in Remark  \ref{remark:3}, with large sample sizes, failure to violate the instrumental variable inequalities shows that an instrumental variable model is compatible with the observed data, but does not validate such a model.  
Specifically, if one believes that the sample size is sufficiently large, then the results in Table \ref{tab:401k} show that  \cite{okui2012doubly}'s instrumental variable model is compatible with the observed data. One should use their model if one also believes that college proximity only affects earnings through education, and  that there is no unmeasured confounding after
adjusting for experience, race and region of residence. In contrast, one should not trust the instrumental variable model conditional only on  experience, regardless of one's prior substantive belief.}

\section{Discussion}
\label{sec:discussion}

Although instrumental variable methods are widely used to identify causal effects in the presence of unmeasured confounding, their assumptions  have mainly been assessed based on subject-matter arguments rather than statistical evidence. However, there are controversies in the validity of many instruments especially if they are not randomized; for example, see \cite{rosenzweig2000natural} for a discussion on using natural experiments as instruments. Thus it should be routine  to check the instrumental variable model against the observed data; see also \cite{didelez2010assumptions}. In this paper, we introduce a simple approach for testing the binary instrumental variable model. 

Our approach can be extended to  test discrete instrumental variable models with binary outcomes. According to \cite{pearl1995causal}, {testable implications in this case include}
\begin{equation}
\label{eqn:iv_discrete}
	\max\{p(0,d\mid 0),  \ldots, p(0,d\mid z_{\text{max}})\} + 	\max\{p(1,d\mid 0), \ldots, p(1,d\mid z_{\text{max}})\} \leq 1 \quad (d = 0, \ldots, d_{\text{max}}),
\end{equation}
{where $Z$ takes value in  $0,\ldots, z_{\text{max}}$ and $D$ takes value in $0, \ldots, d_{\text{max}}$. 
	With slight modifications on the multiplicity adjustments, the techniques introduced in this paper can be used to test inequalities \eqref{eqn:iv_discrete}; see Appendix for details.
	In general, there are other observed data constraints implied by the discrete instrumental variable model \citep{bonet2001instrumentality}, the testing of which is an interesting topic for future research.}

Monotonicity is also often assumed in the instrumental variable analysis. See 
\cite{huber2015testing} for a joint test of the {unconditional} instrumental variable model and the monotonicity assumption.  
It is a future research problem to extend the proposed approach to test the binary instrumental variable model under monotonicity.

Although we have focused primarily on testing the binary instrumental variable model, as we explain in Section \ref{sec:unconditional}, with randomized experiments our proposed tests can be directly applied to identify  the sign of the average controlled directed effects $\textsc{ACDE}(d) = E\{Y(z=1,d=d)\} - E\{Y(z=0,d=d)\} \ (d = 0,1)$. These average controlled direct effects quantify
the extent to which the randomized treatment $Z$ affects the outcome $Y$  not through the mediator $D$, which are important for explaining causal mechanisms.

\section*{Acknowledgement}

{We thank Chengchun Shi for helpful comments. This research was supported by grants from U.S. National Institutes of Health and
Office of Naval Research.}  This work was initiated when the first author was a graduate student at the University of Washington.

\section*{Supplementary material}
{Supplementary material available at \emph{Biometrika} online includes code for the data application.}

\section*{Appendix}

\subsection*{Proof of Theorem \ref{thm:main}}

\begin{proof}
	The second and third claims in Theorem \ref{thm:main}	follow directly from the assumption that the size of $\phi^{dy}$ goes to 0 asymptotically in the interior of the null space defined by $\h^{dy}_0$.
	We now consider the case where two inequalities in \eqref{eqn:test_implication} hold with equality at the true value $\dot\zeta$. Without loss of generality, we assume $\dot{\Delta}^{00} = \dot{\Delta}^{01} = 0,$ where the dot denotes the true value. As $\sum_{d,y} \dot{\Delta}^{dy} = -2$ and $-1 \leq \dot\Delta^{dy} \leq 0, d=0,1, y=0,1,$ we immediately get  that for $d=1,y=0,1,$ $\dot{\Delta}^{dy} = -1$ and hence ${\rm pr}(Q^{dy}=1\mid Z=1) = 0$ and ${\rm pr}(Q^{dy}=1\mid Z=0) = 1.$ As a result, for $d=1,y=0,1,$  {one cannot reject $\h^{dy}_0$, with probability 1}. On the other hand, as at most one of $\h^{00}_0$  and $\h^{01}_0$ can be violated empirically, they  cannot be rejected simultaneously following our assumptions on $\phi^{dy}$. The probability of rejecting at least one of $\phi^{00}$ and $\phi^{01}$ hence equals to $\alpha$ in this case.
	\end{proof}

\subsection*{Multiplicity adjustment with the discrete instrumental variable model}

Constraints in \eqref{eqn:iv_discrete} can be written as
\begin{equation}
\label{eqn:impli}
		p(0,d\mid z_1) + p(1,d \mid z_2 ) \leq 1 \quad (z_1,z_2 = 1,\ldots, z_{\text{max}}, z_1\neq z_2; d=1,\ldots, d_{\text{max}}).
\end{equation}
There are  $d_{\text{max}} z_{\text{max}} (z_{\text{max}}-1)$ inequalities in \eqref{eqn:impli}, the left-hand sides of which sum to $z_{\text{max}} (z_{\text{max}}-1)$. Hence  at most  $z_{\text{max}} (z_{\text{max}}-1)$ of them can hold with equality simultaneously.  Similar to Theorem \ref{thm:main}, the proposed testing procedure for the unconditional discrete instrumental variable model proceeds as follows: reject \eqref{eqn:impli} if  for $d=0,\ldots,d_{\text{max}}, y=0,1,$ at least one of $\h_0^{dy}$  is rejected by $\phi^{dy}$  at level $\alpha/\left\{  z_{\text{max}} (z_{\text{max}}-1)  \right\}$. For the conditional discrete instrumental variable model, the Bonferroni correction is appropriate; see also Remark \ref{remark:2}.

 \thispagestyle{empty}
\bibliographystyle{biometrika}
\bibliography{causal}

\end{document}